\documentstyle[preprint,aps]{revtex}
\newcommand{\sla}[1]{\mbox{$/ \!\! \! \! #1 $}}
\newcommand{\be}{\begin{equation}}
\newcommand{\ee}{\end{equation}}
\newcommand{\bea}{\begin{eqnarray}}
\newcommand{\eea}{\end{eqnarray}}

\begin{document}
\draft
\title{Braaten-Pisarski Method at Finite Chemical Potential
\footnote{supported by BMFT and GSI Darmstadt}}
\author{Hans Vija\footnote{present address:
Physics Department, University of Washington, Seattle, WA 98195, USA}
and Markus H. Thoma}
\address{Institut f\"ur Theoretische Physik, Universit\"at Giessen,\\
35392 Giessen, Germany\footnote{e-mail: vija@theorie.physik.uni-giessen.de,
thoma@theorie.physik.uni-giessen.de}}
\maketitle

\begin {abstract}
The effective perturbation theory developed by Braaten and Pisarski for
gauge theories at finite temperature is extended to finite chemical
potential. As a first application the collisional energy loss of a
heavy quark propagating through a quark-gluon plasma with non-vanishing
quark chemical potential is considered. Assuming  $\mu /T\simeq 1$,
motivated by numerical simulations of heavy ion collisions at RHIC energies,
we find that the effect of the quark chemical potential is rather small,
unless the energy density instead of the temperature is fixed.
\end{abstract}


\newpage

\section{Introduction}
QCD lattice calculations predict a phase transition from hadronic
matter to a phase of deconfined quarks and gluons (QGP)
above a critical energy density $\epsilon_c \approx 1$  GeV/fm$^3$
corresponding to a critical temperature $ T_c \approx 150$ MeV
(see, e.g. \cite{r1}). The creation of a QGP is expected to
happen in ultrarelativistic heavy ion collision and in the early universe
shortly after the big bang. An overview of the present status of the
experiments is given in \cite{r2}.

Depending on the available center of mass energy of the colliding
nuclei, one can think of two scenarios \cite{r3}:
Below an energy per nucleon of approximately $5$ GeV/A complete stopping
of the participating nuclei takes place. It is believed that in this region
of high compression
and high temperature a QGP containing more quarks than anti-quarks
might be produced.
Above an energy of about a few $100$ GeV/A, on the other hand, full
transparency is expected. The QGP is believed to be created in the
region between the two expelling and highly excited nuclei by vacuum
polarization leading to a vanishing baryon number in the fireball.

We consider the  quark chemical potential $\mu $, which is related
to the baryo-chemical potential  $\mu_b$ through $\mu = \mu_b/3$ and
measures the deviation from the balance of quarks and anti-quarks. An equal
amount of quarks and anti-quarks corresponds to $\mu = 0$, whereas $\mu > 0 $
describes an excess of quarks over anti-quarks. It is generally agreed that
$\mu/T \simeq 0$
at LHC energies. At RHIC energies, however,  there might still exist a
considerable amount of stopping leading to $\mu /T\simeq 1$ - 2 \cite{r4}
as predicted by numerical simulations (RQMD) \cite{r5}.

Properties of a QGP at finite chemical potential cannot be described by
lattice calculations so far. Perturbative QCD at finite temperature and
chemical potential suffers from infrared singularities and gauge dependent
results. These problems can be avoided for the most physical quantities of
interest by adopting an effective perturbation theory developed by Braaten
and Pisarski, which is based on the use of resumed Green functions instead of
bare ones \cite{r6}. In this way consistent results, i.e. complete to
leading order in the coupling constant and gauge independent, can be derived
in the weak coupling limit, $g\ll 1$.

Here we will consider the Braaten-Pisarski method at finite temperature as
well as chemical potential, which is relevant for ultrarelativistic
heavy ion collisions if no complete transparency occurs. Up to now only
the cases $T\neq 0$, $\mu =0$ \cite{r6} and $T=0$, $\mu \neq 0$ in
an astrophysical context \cite{r7} have been discussed.

The next section treats the technical aspects of introducing the chemical
potential into thermal field theory. We extend the  Braaten-Pisarski method
to finite chemical potential and apply the results to the calculation
of the energy loss of a heavy quark in a QGP as a typical example
in the third section. Throughout the paper
we use the metric tensor $g = (+ - - -)$, natural units
($\hbar = c = k = 1$) and the notation $K=(k_0, {\bf k})$ and $k=|{\bf k}|$
for the momenta.

\section{Perturbative techniques at finite chemical potential}

\subsection{Saclay method}

The Braaten-Pisarski method starts from the imaginary time formalism
where one has to sum over discrete energies (Matsubara frequencies).
For this purpose the Saclay method was developed in the sixties by Balian
and De Dominics \cite{r8} and used by Pisarski in \cite{r9} allowing
an elegant way to evaluate finite temperature loops.
The idea is based upon the 'mixed` representation of the propagators, i.e.
a momentum representation of space components and a space representation
of the time component. In the case of the bare fermion propagator including
the chemical potential and neglecting the bare mass it reads:
\bea
S(K)  & = &  \frac{1}{\sla{K}} = \sla{K}\> \widetilde{\Delta}(K),\nonumber\\
\widetilde{\Delta}(K) & \equiv & \frac{1}{K^2} = -\int^\beta_0  d\tau \>
e^{k_0\tau}\> \widetilde{\Delta}(\tau,k),\nonumber\\
\widetilde{\Delta}(\tau,k) & = & -T \> \sum^\infty_{n=-\infty} e^{-k_0\tau}
\> \frac{1}{k^2_0-k^2}, \nonumber\\
k_0 &=& (2n+1)\> i\> \pi \> T + \mu.
\label{e1}
\eea
The sum is evaluated best by a complex integration \cite{r10}
yielding the Saclay representation of the fermion propagator:
\be
\widetilde{\Delta}(\tau ,k) = \frac {1}{2k}\> \left \{e^{-k\tau} \>
\left[1-n_F(k-\mu)\right] - e^{k\tau}\> n_F(k+\mu)\right\},
\label{e2}
\ee
where $n_F(k\pm \mu) = 1/[\exp((k\pm \mu)/T)+1]$ denotes the Fermi
distribution function. The advantage of this representation is the easy
summation over the Matsubara frequencies since the discrete energy $k_0$
appears only in $\exp (k_0\tau )$ in the propagator $\widetilde \Delta (K)$ .
The derivation of the boson propagator containing no chemical potential
proceeds analogously \cite{r9}.

\subsection{Hard Thermal Loops}

Following the program given by the Braaten-Pisarski resummation
technique in ref.\cite{r6} we calculate the Hard Thermal Loops (HTL)
at finite chemical potential, first. HTL are one-loop diagrams for
which only the contribution from hard loop momenta of the order $T$ or
larger is considered. This approximation is equivalent to the leading
term of the high temperature expansion of the diagram under consideration.
Here, we attach importance mainly to changes introduced by the chemical
potential and refer to the literature otherwise.

First we discuss the HTL approximation of the gluon self energy, including
the quark chemical potential and using the Saclay method. The only
modification
to the HTL polarization tensor compared to the zero chemical potential case
comes from the diagram of fig.1. Using standard Feynman rules in the
imaginary time formalism and the Saclay representation of the
propagators, where we consider one fermion line as a quark with
chemical potential $\mu $ and the other one as an anti-quark with $-\mu $,
we obtain analogously to the case of zero chemical potential \cite{r6},
assuming the HTL limit,
\bea
\Pi _L (K) & \equiv & \Pi _{00} (K)\nonumber \\
& = & -3\> m_g^2\> \left (1-\frac {k_0}{2k}\> \ln \frac {k_0+k}{k_0-k}
\right )
\label{e3}
\eea
for the longitudinal part and
\bea
\Pi _T (K) & \equiv & \frac {1}{2}\> \left (\delta _{ij}-\frac {k_ik_j}{k^2}
\right )\> \Pi _{ij} \nonumber \\
& = & \frac {3}{2}\> m_g^2\> \frac {k_0^2}{k^2}\> \left [1-\left (1-\frac
{k^2}{k_0^2}\right )\> \frac {k_0}{2k}\> \ln \frac {k_0+k}{k_0-k} \right ]
\label{e4}
\eea
for the transverse part of the gluon polarization tensor. Here
\bea
m_g^2 & = & \frac {g^2T^2}{3}\> \left (1+\frac {\tilde N_f}{6}\right ),
\nonumber \\
\tilde N_f & = & N_f+\frac {3}{\pi ^2}\> \sum _f \frac {\mu _f^2}{T^2},
\label{e5}
\eea
is called the effective gluon mass, where $N_f$ denotes the number of thermal
flavors in the QGP and $\mu _f$ the corresponding chemical potentials. In
the following a QGP consisting of up and down quarks with $\mu _u=\mu _d
=\mu $ and all other $\mu _f=0$ is assumed. The only change in the gluon
self energy due to the introduction of a chemical potential is given
by the effective gluon mass which depends now on the chemical potential
in accordance with the result of the high temperature limit \cite{r11}.
In the limit $T\rightarrow 0$ the result (\ref{e3}) to (\ref{e5}) also
reduces to the one found in ref.\cite{r7}.

The same behavior is observed for the HTL quark self energy. One
simply has to replace the effective quark mass by the chemical
potential dependent one
\be
m_q^2=\frac {g^2T^2}{6}\> \left (1+\frac {1}{\pi ^2}\> \frac {\mu ^2}{T^2}
\right )
\label{e6}
\ee
already quoted in ref.\cite{r12}. (In the course of the derivation of
the HTL quark self energy we had to assume that the chemical potential
is not much larger than the temperature, $\mu \ll T/g$, whereas this
restriction was not necessary for the gluon self energy.)

Owing to Ward identities the HTL self energies also imply the existence of
HTL vertex corrections \cite{r6}. They also will be modified in the presence
of a finite chemical potential in a similar way as the self energies. However,
for the most applications considered so far they can be neglected.

\subsection{Effective propagators}

The effective gluon and quark propagators follow by resuming the HTL self
energies by means of a Dyson-Schwinger equation \cite{r6}. Hence they
differ from the ones at zero chemical potential (see, e.g. ref.\cite{r23}
for the gluon propagator and ref.\cite{r33} for the quark propagator)
only by the effective masses (\ref{e5}) and (\ref{e6}). According to
the Braaten-Pisarski method effective Green functions
(propagators or vertices) have to be used only if all legs of the Green
function have soft momenta of the order $gT$. Otherwise the use of bare
Green functions is sufficient. In this way gauge independent results,
which are complete to leading order in the coupling constant, are obtained
for physical quantities. Moreover, due to the self energies in the
denominator of the effective propagators important medium effects such
as Debye screening, which improves the infrared behavior, and
Landau damping are included. Thus the Braaten-Pisarski method means a
crucial improvement compared to the naive perturbation theory based
on bare Green functions. However, the application of the Braaten-Pisarski
method is limited by two problems \cite{r13}.
First, the Braaten-Pisarski method
relies on the distinction between hard and soft momenta, i.e. on the
weak coupling limit assumption, $g\ll 1$. The validity of the extrapolation
of the results to realistic values of the coupling constant, $g=1.5$ - 2.5,
in ultrarelativistic heavy ion collisions, has to be checked for each
quantity individually. In general it appears to work for
particles with large momenta, $p\gg T$. Secondly, the effective
gluon propagator suffers from the absence of a static magnetic screening mass
in the HTL approximation of the gluon polarization tensor, thus still
leading to infrared singularities in certain quantities, as e.g. the damping
rate of a hard parton \cite{r14}. However, so far only a few observables,
e.g. the color relaxation \cite{r15}, are known to be affected by this
problem. The most physical quantities of interest, such as the collisional
energy loss of energetic partons \cite{r16}, transport rates of quarks and
gluons \cite{r17}, the viscosity of the QGP \cite{r18}, and the production
rate of hard photons in the QGP \cite{r19}, are to leading order infrared
finite by using effective propagators due to dynamical magnetic screening
\cite{r20}.

Summarizing this section, the generalization of the Braaten-Pisarski method
to finite chemical potential is straightforward by modifying the effective
parton masses, containing now also the chemical potential, that enter
into the effective propagators.

\section{Energy loss of a heavy quark}

The energy loss of energetic partons in a QGP is related to the
phenomenon of jet quenching in ultrarelativistic heavy ion collisions
\cite{r21}, which might serve as a signature for the QGP formation \cite{r22}.
The energy loss of a parton in a QGP is caused by elastic scattering off
the thermal partons (collisional energy loss) and gluon bremsstrahlung
(radiative energy loss). The collisional contribution follows from the
damping rate by introducing the energy transfer per parton collision
as a weight under the integral defining the damping rate \cite{r23}.

The damping rate of a hard quark follows from the imaginary part of the
quark self energy
diagram shown in fig.2, which comes from the imaginary part
of the effective gluon propagator corresponding to Landau damping of
a virtual collective gluon mode. This damping mechanism is equivalent
to elastic scattering off the thermal partons via the exchange of a
collective gluon (plasma wave), as can be seen from cutting the diagram
in fig.2. As mentioned above, the damping rate, which is quadratic
infrared divergent using
bare propagators, turns out to be logarithmically infrared divergent
applying the Braaten-Pisarski method reflecting the absence of static magnetic
screening in the effective gluon propagator. Integrating over the loop
momentum a cancellation of the
effective gluon mass takes place \cite{r17}. Consequently the damping rate
of a hard quark, given to logarithmic accuracy by $\gamma _q=g^2T/(3\pi )\,
\ln (1/g)$ \cite{r24}, does not depend on the chemical potential.
(The Fermi-Dirac
distributions from the internal quark line is exponentially suppressed
for hard quarks \cite{r23}.)

The additional energy transfer
factor in the loop integral reduces the infrared divergence
and leads to a finite result using the effective gluon propagator at
soft momentum transfer, $q{\buildrel <\over \sim}gT$. For large momentum
transfer, $q{\buildrel >\over \sim}T$, the exchange of a
bare gluon in the elastic scattering diagram is sufficient. According
to the prescription by Braaten and Yuan \cite{r25} the both contributions
can be matched onto each other by assuming a separation scale,
$gT\ll q^\star \ll T$, in the weak coupling limit. Adding up the soft and
hard contributions, the separation scale drops out of the final result.
In this way a infrared finite and gauge independent result for the
collisional energy loss of a heavy quark is found, which is complete to
leading order of the coupling constant and independent of arbitrary
parameters such as infrared and ultraviolet momentum cut-offs or
separation scales.

Taking a finite quark chemical potential into account, the derivation of
the soft part proceeds analogously to the case of vanishing chemical
potential \cite{r23,r16}. The only change comes from the effective gluon mass
that does not cancel as opposed to the case of
the damping rate. The hard part requires
integrations over Fermi-Dirac distributions, which can be achieved only
numerically at finite chemical potential. It involves integrals of the
form
\bea
\int^\infty_0 dk\> [n_F(k-\mu)+n_F(k+\mu)] \>k & = &
\frac{\pi^2 T^2}{6}\> \left(1+\frac{3\mu^2}{\pi^2 T^2}\right),\nonumber \\
\int^\infty_0 dk\> [n_F(k-\mu)+n_F(k+\mu)] \frac{k}{q^\star} & = &
\frac{\pi^2 T^2}{6}\> \left(1+\frac{3\mu^2}{\pi^2 T^2}\right)\nonumber \\
&& \left[\ln\frac{T}{q^\star}+\ln 2+1-\gamma+
\frac{\zeta'(2)}{\zeta(2)}+F\left(\frac{\mu}{t}\right)\right],
\label{e7}
\eea
where $\gamma = 0.57722...$ is Euler's constant, $\zeta(z)$ Riemann's
zeta function with $\zeta'(2)/\zeta(2)=-0.56996...$, and the function
$F(\mu /T)$ with $F(0)=0$, shown in fig.3, has to be determined numerically.

The final result for the energy loss of a heavy quark with mass $M\gg T$,
energy $E$, and velocity $v$ in a QGP of $N_f$ thermal flavors
after adding up the soft and hard contributions
is given by rather simple expressions in the limits $E\ll M^2/T$,
\bea
-\frac {dE}{dx} & = & \frac {8\pi \alpha _s^2T^2}{3}\> \left (1+\frac
{\tilde N_f}{6}\right )\> \left (\frac {1}{v}-\frac {1-v^2}{2v^2}\>
\ln \frac
{1+v}{1-v}\right )\nonumber \\
&&\left [\ln \left (2^{\frac {\tilde N_f}{6+\tilde N_f}}
\> B(v)\> \frac {ET}{m_gM}\right ) + \frac {\tilde N_f}
{6+\tilde N_f}\> F\left (\frac {\mu }{T}\right )\right ],
\label{e8}
\eea
and $E\gg M^2/T$,
\be
-\frac {dE}{dx} = \frac {8\pi \alpha _s^2T^2}{3}\> \left (1+\frac {\tilde
N_f}{6}\right )\> \left [\ln \left (2^{\frac {\tilde N_f}
{2(6+\tilde N_f)}}\> 0.920\> \frac {\sqrt{ET}}{m_g}\right ) +
\frac {\tilde N_f}{2(6+\tilde N_f)}\>
F\left (\frac {\mu }{T}\right )\right ]
\label{e9}
\ee
respectively, where
$B(v)$ is a smooth function of the velocity that increases monotonically
from $B(0)=0.604$ to a maximum of $B(0.88)=0.731$, and then decreases
to $B(1)=0.629$ \cite{r16}. The expressions (\ref{e8}) and (\ref{e9})
extend the Bethe-Bloch formula \cite{r26} to the case of a relativistic
plasma with finite temperature and chemical potential. They differ from
the corresponding expressions at zero chemical potential \cite{r16} by the
effective gluon mass (\ref{e5}), the effective flavor number $\tilde N_f$,
and the additional function $F(\mu /T)$. The details of the
derivation of (\ref{e8}) and (\ref{e9}) are presented in ref.\cite{r27}.

Figs.4 and 5 show the energy loss of a charm and bottom quark
extrapolated to $\alpha _s=0.2$ at $T=250$ MeV and $N_f=2$
for $\mu /T=0$, 1, and 2 as a function of the momentum $p$. In the case
of a charm quark (\ref{e8}) is matched onto (\ref{e9}) at some intermediate
momentum by demanding $dE/dx$ to be continuous, whereas for a bottom
quark (\ref{e8}) is sufficient for the momentum range chosen in fig.5. At
momenta below a few GeV the energy loss turns out to be negative
reflecting the breakdown of the extrapolation to realistic values
of the coupling constant for small momenta \cite{r13,r16}. The enhancement
of $dE/dx$ with increasing chemical potential is rather small (below
about 25\% at $\mu /T=2$) since there is a partial cancellation
between the factor $\tilde N_f$ increasing with $\mu $ and the function
$F(\mu /T)$ which is negative for $\mu >0$ (fig.3). This demonstrates
that the temperature is more important than the chemical potential for the
energy loss.

However, if one fixes the energy density, given e.g. by the MIT-bag
equation of state \cite{r28}
\be
\epsilon (\mu ,T) = \left (\frac{37\pi^2}{30}-\frac{11\pi\alpha_s}{3}\right )
\> T^4 + 3\left(1-\frac{2\alpha_s}{\pi}\right)\> T^2\> \mu ^2 +
\frac{3}{2\pi^2}\left(1-\frac{2\alpha _s }{\pi}\right )\> \mu ^4 + B
\label{e10}
\ee
with the bag constant $B=(200$ MeV$^4)$, increasing the chemical potential
corresponds to a decrease of the temperature resulting in a significant
reduction of the energy loss as shown in fig.6. (At $\mu =0$ a temperature
of $T=250$ MeV corresponds to $\epsilon = 5.232$ GeV/fm$^3$.) Considering
ultrarelativistic heavy ion collisions the choice of a certain energy
density instead of a temperature might be reasonable as the former is
closer related to the collision energy \cite{r29}.

The collisional
energy loss of a light quark follows from (\ref{e9}) by replacing
the energy $E$ by a maximum momentum transfer $\tilde q\simeq E/2$,
the energy loss of a gluon by multiplying the one of a light quark by
a color factor of $9/4$ \cite{r30}.
The radiative energy loss by gluon bremsstrahlung
is much more complicated and has not been addressed by the Braaten-Pisarski
method so far. Estimates based on naive perturbation theory indicate
the dominance of the radiative energy loss over the collisional one
in a QGP \cite{r31}.

\section{Conclusions}

In the present work the Braaten-Pisarski method has been extended to
finite chemical potential. The effective propagators, based on a
resummation of the HTL self energies, differ from the one at zero
chemical potential only by the effective masses which now depend on
the chemical potential. The Braaten-Pisarski method can be used for
calculating consistently quantities of the QGP that are logarithmically
infrared divergent if bare Green Functions are utilized.
Interesting observables related to the evolution and detection of a QGP
in ultrarelativistic heavy ion collisions, belonging to this class of
quantities, are the energy loss of energetic partons propagating through
the QGP, the transport rates of quarks and gluons, the viscosity of
the QGP, and the production rate of hard photons in the QGP.

Here the collisional energy loss of a heavy quark, representing a typical
example for the application of the Braaten-Pisarski method, has been
computed to leading order at finite temperature as well as finite quark
chemical potential. This might be relevant for heavy ion collisions
at RHIC energies, for which numerical simulations (RQMD) predict a
considerable amount of stopping leading to a quark chemical potential
of the order of $\mu \simeq 1$ - 2 $T$. The energy loss of a heavy quark
has been calculated consistently adopting the Braaten-Yuan prescription,
which corresponds to a decomposition in a soft contribution, calculable
with an effective gluon propagator, and a hard, within naive perturbation
theory accessible contribution. Besides the modification caused by
the effective gluon mass (\ref{e5})
in the soft part, momentum integrations over
the Fermi-Dirac distribution functions in the hard part
lead to an additional term (see (\ref{e7})),
which requires a numerical evaluation at finite
chemical potential. However, these changes are small,
typical below 25\%, for $\mu
\leq 2T$ as long as the temperature is fixed, showing the dominant role
of the temperature over the chemical potential. If, on the other hand, a
given energy density $\epsilon (\mu ,T)$ is assumed, corresponding to a
decrease of the temperature with increasing chemical potential, a
significant suppression is observed.

The same conclusions apply to the transport rates of quarks and gluons,
which determine thermalization times and the viscosity of the QGP \cite{r17},
The transport rate follows from the damping rate by including a transport
weight describing momentum relaxation. Owing to this transport weight the
computation of the transport rate proceeds similarly to the one of the energy
loss via the Braaten-Yuan prescription. The details of this calculation can
be found in ref.\cite{r27}.

The production rate of hard photons from a QGP can be derived to leading
order again by using the Braaten-Yuan prescription, where the soft part
follows from the photon self energy containing one effective quark
propagator and the hard part from the tree matrix elements of Compton
scattering and pair annihilation involving one gluon and one photon
\cite{r19}. The calculation of the hard photon production rate
at finite quark chemical potential considering the effective quark propagator
modified by the effective quark mass (\ref{e6}) will be presented in a
in a forthcoming publication\cite{r34}.

\acknowledgements
We would like to thank T.S. Bir\'o for valuable discussions.

\begin{figure}
\caption{Part of the HTL gluon polarization tensor containing the quark
chemical potential}
\end{figure}

\begin{figure}
\caption{Quark self energy containing the effective gluon propagator}
\end{figure}

\begin{figure}
\caption{The function $F(\mu /T)$}
\end{figure}

\begin{figure}
\caption{Energy loss of a charm quark for $\mu /T=0$, 1 and 2 for a given
temperature}
\end{figure}

\begin{figure}
\caption{Energy loss of a bottom quark for $\mu /T=0$, 1 and 2 for a given
temperature}
\end{figure}

\begin{figure}
\caption{Energy loss of a bottom quark for $\mu /T=2$ for a given energy
density compared to a given temperature}
\end{figure}

\end{document}